\documentclass[12pt]{article}

\usepackage{graphicx}
\usepackage{amsmath}
\usepackage{amsfonts}
\usepackage{amssymb}

\def\la{\mathrel{\mathpalette\fun <}}
\def\ga{\mathrel{\mathpalette\fun >}}
\def\fun#1#2{\lower3.6pt\vbox{\baselineskip0pt\lineskip.9pt
\ialign{$\mathsurround=0pt#1\hfil##\hfil$\crcr#2\crcr\sim\crcr}}}

\begin{document}

\null
\vskip 2cm

\begin{center}
{\bf\large Modification of Coulomb law and energy levels \\ \smallskip
of the hydrogen atom in a superstrong magnetic field}
\end{center}

\bigskip
\centerline{B.~Machet
\footnote{machet@lpthe.jussieu.fr}}
\centerline{\small{\em LPTHE
\footnote{Laboratoire de Physique Th\'eorique et Hautes \'Energies, Unit\'e
Mixte de Recherche UMR 7589 (CNRS / UPMC Univ Paris 06)
, tour 13-14, 4\raise 3pt \hbox{\tiny \`eme} \'etage,
          UPMC Univ Paris 06, BP 126, 4 place Jussieu,
          F-75252 Paris Cedex 05 (France)}
, UMR 7589 (CNRS \ UPMC Univ Paris 06), Paris}}
\smallskip
\centerline{M.~I.~Vysotsky
\footnote{vysotsky@itep.ru}}
\centerline{\small{\em ITEP
\footnote{ITEP,
25 Bolshaya Cheremushkinskaya Ul., RU-117259 Moscow (Russia)}
, Moscow}}

\medskip
\centerline{\small{\em 26 November 2010. Revised 17 December 2010}}

\vskip 1cm

We obtain the following analytical formula which describes the
dependence of the electric potential of a point-like charge on the
distance away from it in the direction of an external magnetic
field $B$:

${\bf\Phi}(z) = e/|z|\left[ 1- {\rm exp}(-\sqrt{6m_e^2}\;|z|) + {\rm
exp}(-\sqrt{(2/\pi) e^3 B + 6m_e^2}\;|z|)\right]$. \\ The deviation
from Coulomb's law becomes essential for $B > 3 \pi B_{cr}/\alpha
= 3 \pi m_e^2/e^3 \approx 6 \cdot 10^{16}$ G. In such superstrong 
fields, electrons are
ultra-relativistic except those which occupy the lowest Landau
level (LLL) and which have the energy $\varepsilon_0^2 = m_e^2 +
p_z^2$. The energy spectrum on which LLL splits in the presence of
the atomic nucleus is found analytically. For $B > 3 \pi
B_{cr}/\alpha$ it substantially differs from the one obtained
without accounting for the modification of the atomic potential.

\vskip .5 cm

%


\section{Introduction}

In the pioneering papers \cite{1} an exponential modification of
Coulomb's law by superstrong magnetic fields $B > m_e^2/e^3 = 137
B_{cr}$ was discovered ($B_{cr} \equiv m_e^2/e = 4.4 \cdot 10^{13} G = 4.4 \cdot
10^{9} T$ is the so-called {\it critical} or {\it Schwinger}
 magnetic field \footnote{All formulas are written in Gauss units,
where $\alpha = e^2 = 1/137$ ($\hbar=c=1$ is implied). }).
It originates from the strong (linear) dependence of
the photon polarization operator on the external $B$.
The Coulomb potential gets modified at distances $1/m_e > r >
1/\sqrt{e^3 B} \equiv a_H/e$, where $a_H$ is the Landau radius
(the size of the ground state electron wave function in
the direction transverse to the homogeneous magnetic field).
In \cite{1} the shape of the modified potential
was determined numerically.

As found long ago in papers \cite{2} the photon polarization
operator in a strong magnetic fields $B > B_{cr}$
is dominated by electron and positron states belonging to 
the lowest Landau level (LLL) and it
 factorizes
into transverse and longitudinal (with respect to the direction
of $B$) parts. Its
dependence on  the longitudinal and time-like components of the
momentum coincides  with that in two-dimensional QED (with no
$B$). In paper
\cite{3} a simple interpolating formula for the photon polarization
operator in two-dimensional QED was found, which allows to find an
analytical expression for the potential of a point-like electric
charge in $D = 2$ QED. For light ``electrons'' propagating in the
loop ($m < g$, where $g$ is the electric charge in $D = 2$ QED),
the electric potential gets screened  at all distances larger
than $1/(2g)$.

According to \cite{3} the deviation of the interpolating formula
from the exact expression for the polarization operator in $D = 2$
does not exceed 10\%. In Section 2 we will show that
 for various values of the ratio
$g/m$ the accuracy of the analytical
formula for the potential in $D=2$ is always better than
4\% for all distances $z$.

For the realistic case $D=4$ only the asymptotics of the electric
potential at large ($|z| \gg 1/m_e$) and small ($|z| \ll 1/m_e$)
distances were found in \cite{3}. In agreement with \cite{1} the
screening only takes place at short distances $|z| \ll 1/m_e$. In
Section 3, using the interpolating formula for the polarization operator
of \cite{3}, we
will obtain an analytical expression for the dependence of the
electric potential of a point-like charge on the distance to this
charge in the longitudinal direction (parallel to the magnetic field $B$)
and at zero transverse direction. This potential determines
the energies of atomic electrons. The spectrum of electrons
originating from LLL
will be found in Section 4.

In \cite{1} the ground state energy of a hydrogen atom in a
superstrong magnetic field was found in the shallow-well
approximation. The detailed analytical study and comparison with
numerical results for the atomic levels in an external magnetic
field performed in paper \cite{4} clearly demonstrates that
the shallow-well approximation used originally for this problem in
textbook \cite{5} has a very poor accuracy (better to say no
accuracy at all). An algebraic expression for the energies of the
ground and excited states was obtained in \cite{4} which
reproduces the results of numerical calculations with high
accuracy. In Section 4 an analogous expression, but which now takes
screening into account will be derived. It yields a value
of the ground state energy in the limit of infinite $B$ which is
$E_0 = -1.7$ keV, which strongly differs from the result
of the shallow-well approximation \cite{1}: $E_0^{sw} = -4.0$ keV.

Excited states can be classified with respect to the
change of sign of the $z$ coordinate and divided
into even and odd states. Odd states in the magnetic field
$B\ga m_e^2/e^3 \equiv B_{cr}/\alpha$ follow unperturbed
Coulomb levels with very high
accuracy: $E_{odd} = -me^4/(2n^2)$, $n = 1,2,...$ and the
screening does not alter this result. Concerning even states, at
finite $B$ they considerably deviate from Coulomb levels. These
deviations decrease with increasing $B$ like
$1/\log(B)$, and, in the limit of infinite $B$, the even and odd states
become degenerate \cite{6}. The screening changes this result
qualitatively and lifts this degeneracy: the energies of even states
remain considerably higher
than those of corresponding odd states, even in the limit $B \to\infty$.

The dependence of the electron mass on the external magnetic field is
discussed in Appendix. 

We summarize our results
in the Conclusion.

\section{Coulomb potential modification in $\boldsymbol{D = 2}$ QED}

The fermionic part of the $D=2$ QED Lagrangian 
is $L=\bar{\psi}(i\hat \partial - g \hat A - m)\psi$, 
where $\gamma_0 = \sigma_1, \gamma_1 = i\sigma_2$ are Pauli matrices.
Inserting the  photon polarization operator $\Pi $ into the photon
propagator $D$ leads to the following equations (see Fig. 1).
The tree level Coulomb potential
\begin{equation}
\Phi(\vec k) \equiv A_0(\vec k) = \frac{4\pi g}{\vec k^2} \;
\;\; \label{1001}
\end{equation}
gets transformed into the geometrical series
\begin{equation}
{\bf\Phi} \equiv {\bf A}_0 = D_{00} + D_{00}\Pi_{00}D_{00} +
... \; , \label{1}
\end{equation}
and summing it we obtain
\begin{equation} {\bf\Phi}(k) = -\frac{4\pi g}{k^2
+ \Pi(k^2)} \; , \;\; \Pi_{\mu\nu}
\equiv\left(g_{\mu\nu}-\frac{k_\mu k_\nu}{k^2}\right)\Pi(k^2) \;\;
, \label{2}
\end{equation}
where the expression for the photon polarization operator in $D=2$
should be used. Instead of calculating the fermion loop we can take the
expression for $\Pi$ obtained in the dimensional regularization
method \cite{AB}, substitute $D=2$ in it and divide it by two,
because in two dimensions the traces of $\gamma$-matrices are
proportional to $2$ instead of $4$: 
\begin{equation}
\Pi(k^2) = 4g^2\left[\frac{1}{\sqrt{t(1+t)}}\ln(\sqrt{1+t} +\sqrt
t) -1\right] \equiv -4g^2 P(t) \;\;  \label{3}
\end{equation}
where $t \equiv -k^2/4m^2$ .

This approach resembles the one used in the calculation of
the Uehling-Serber corrections to the
 Coulomb potential in \cite{66}. However we resum the whole geometric series
instead of considering 1-loop correction.

\vbox{
\begin{center}
\bigskip
\includegraphics[width=.8\textwidth]{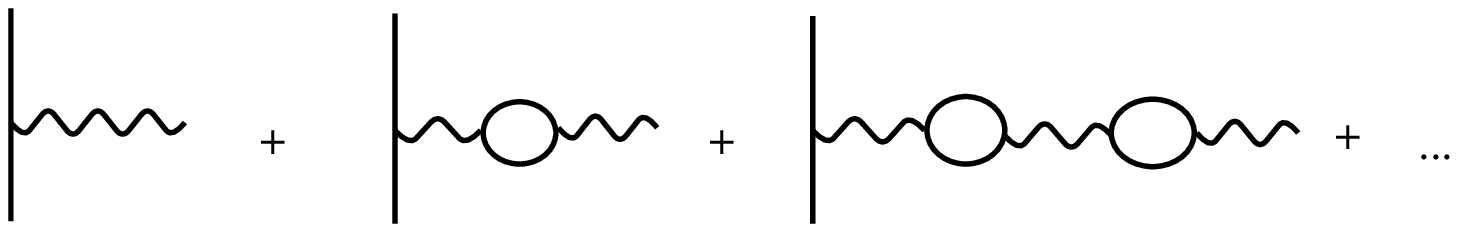}

\medskip
Fig. 1. {\em Modification of the Coulomb potential due to the dressing of the photon propagator.}

\end{center}
}

To obtain the electric potential ${\bf\Phi}(z)$ of a charge at rest we
should set $k_\mu \equiv (k_0, k_\parallel) = (0, k_\parallel)$
(the notation $k_\parallel$ is convenient in the $D=4$ case) and, then,
to
make the Fourier transformation:
\begin{equation}
{\bf\Phi}(z) = 4\pi g \int\limits^\infty_{-\infty} \frac{e^{i
k_\parallel z} dk_\parallel/2\pi}{k_\parallel^2 + 4g^2
P(k_\parallel^2 /4m^2)} \;\; . \label{4}
\end{equation}

Finally, the potential energy of the charges $+g$ and $-g$ is
\begin{equation}
V(z) = -g{\bf\Phi}(z) \;\; . \label{5}
\end{equation}

We did not succeed to analytically perform the integration in
(\ref{4}) with  $P(t)$ given by its exact expression (\ref{3});
instead, an interpolating formula for $P(t)$ was
suggested in \cite{3}. The asymptotics of $P(t)$ are:
\begin{equation}
P(t) = \left\{
\begin{array}{lcl}
\frac{2}{3} t & , & t\ll 1 \\
1 & , & t\gg 1 \;\; ,
\end{array}
\right. \label{6}
\end{equation}
such that the following interpolating formula
\begin{equation}
\overline{P}(t) = \frac{2t}{3+2t}  \label{7}
\end{equation}
has a correct behavior at small and large $t$. Substituting
(\ref{7}) in (\ref{4}) and performing analytically  the integration
yields \cite{3} \footnote{We regularize the infrared divergence of
the $\frac{1}{k_\parallel^2}$ integral making a subtraction at $z=0$.}:
\begin{eqnarray}
{\bf\Phi}(z) & = & 4\pi g\int\limits^{\infty}_{-\infty} \frac{e^{i
k_\parallel z} d k_\parallel/2\pi}{k_\parallel^2 +
4g^2(k_\parallel^2/2m^2)/(3+k_\parallel^2/2m^2)} = \nonumber
\\
& = & \frac{4\pi g}{1+ 2g^2/3m^2}
\int\limits_{-\infty}^{\infty}\left[\frac{1}{k_\parallel^2} +
\frac{2g^2/3m^2}{k_\parallel^2 + 6m^2 + 4g^2}\right]
e^{ik_\parallel z} \frac{dk_\parallel}{2\pi} = \nonumber \\
&=& \frac{4\pi g}{1+ 2g^2/3m^2}\left[-\frac{1}{2}|z| +
\frac{g^2/3m^2}{\sqrt{6m^2 + 4g^2}} {\rm exp}(-\sqrt{6m^2
+4g^2}\;|z|)\right] \;\; .  \label{8}
\end{eqnarray}

For heavy fermions ($m\gg g$) the potential is given by the tree
level expression; radiative corrections are small:
\begin{equation}
{\bf\Phi}(z)\left|
\begin{array}{l}
~~  \\
m \gg g
\end{array}
\right. = -2\pi g|z|\left(1+O\left(\frac{g^2}{m^2}\right)\right)
\;\; . \label{9}
\end{equation}

In the case of light fermions ($m\ll g$) we get a 
much more interesting behavior:
\begin{equation}
{\bf\Phi}(z)\left|
\begin{array}{l}
~~  \\
m \ll g
\end{array}
\right. = \left\{
\begin{array}{lcl}
\pi e^{-2g|z|} & , & z \ll \frac{1}{g} \ln\left(\frac{g}{m}\right) \\
-2\pi g\left(\frac{3m^2}{2g^2}\right)|z| & , & z \gg \frac{1}{g}
\ln\left(\frac{g}{m}\right) \;\; .
\end{array}
\right. \label{10}
\end{equation}

At short distances the potential has a Yukawa behavior,
corresponding to the exchange of a massive vector particle. The fact that
the photon remains massless for nonzero $m$ follows from the behavior
of the potential at large distances: a linear behavior corresponds
to a massless exchange. However, the coupling constant differs from
that at  tree level by the small factor ($3m^2/2g^2$): a screening
of the tree level potential occurs at  $|z|\ga 1/(2g)$. Finally,
for $m=0$  at all $z$ the potential is given by the first line of (\ref{10}):
the photon gets a mass $m_{\gamma} = 2g = 2 \sqrt{\alpha} = e/\sqrt{\pi}$
(where, and only in this formula, $e$ is the coupling constant in
the Heaviside units usually used in Quantum Field Theory)
-- a well-known  result in $D=2$
QED with massless ``electrons'' (Schwinger model), where
a nonzero mass for the gauge boson coexists with
gauge invariance \cite{Schwinger}\cite{ZJ}.

We now study the accuracy of the analytical formula for the
potential (\ref{8}) compared with the  numerical
calculation of the integral (\ref{4}) in  which the exact
expression (\ref{3}) for $P(t)$
is used. In order to normalize the potential
universally for the different values of the ratio $g/m$
we choose an arbitrary
constant up to which the potential is defined in such a way that
${\bf\Phi}(z)=0$ at $z=0$. In Fig. 2 the exact
($ V(z)$) and approximate ($ \bar V(z)$)
potential energies are plotted
together with their asymptotics for $g=0.5$ and $m=0.1$.
The straight line which corresponds to the potential asymptotic
at $z\rightarrow \infty$ crosses the vertical axis at $V(0) \approx
\pi g (1-(3m/2g)^2)\approx 1.42$.

\vbox{
\begin{center}
\bigskip
\includegraphics[width=.7\textwidth]{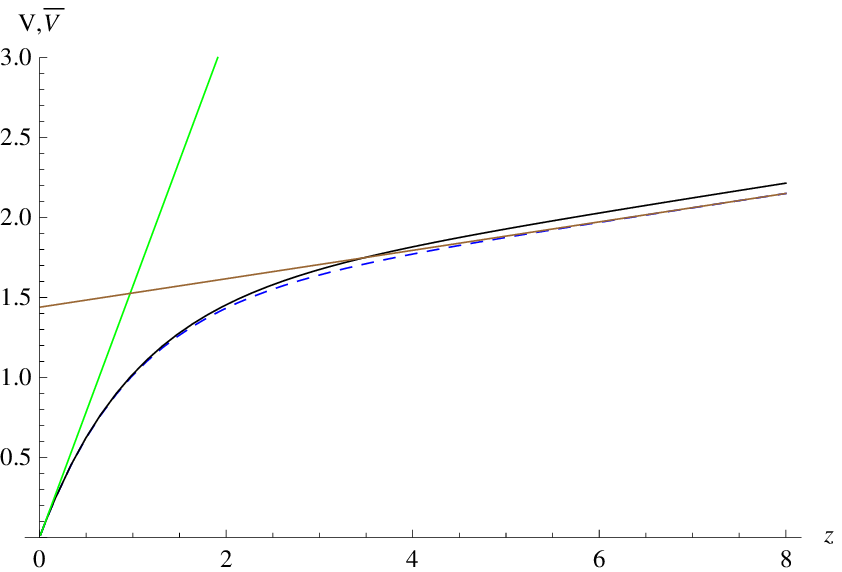}

\medskip
Fig. 2. {\em Potential energies of the charges $+g$ and $-g$ in D=2 for
$g=0.5$, $m=0.1$.
The black curve corresponds to  $P$, the blue-dashed curve -
to $\bar P$.}
\end{center}
}

We see that the accuracy of the interpolating formula is very good. In
Fig. 3 the relative difference $(V - \bar V)/V  $ for the same values
of $g$ and $m$ is shown.

\vbox{
\begin{center}
\bigskip
\includegraphics[width=.7\textwidth]{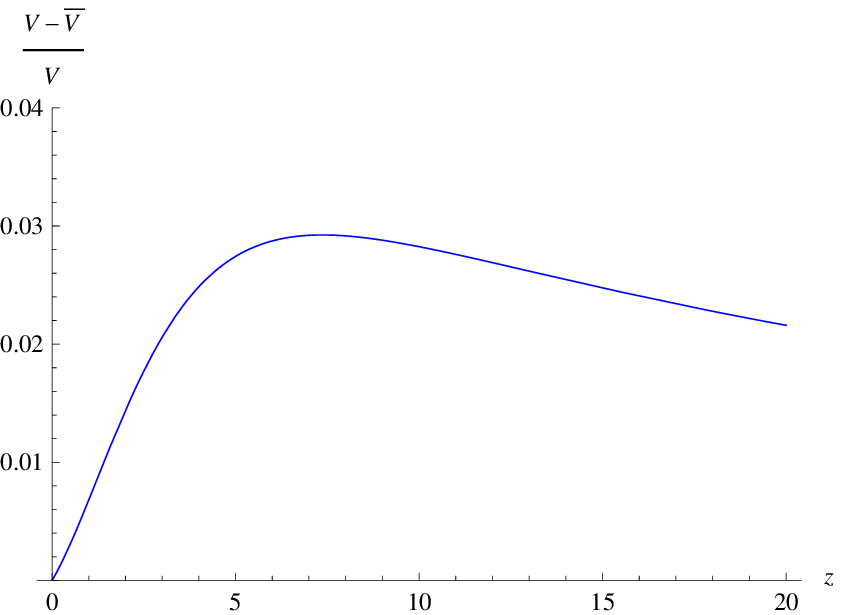}

\medskip
Fig. 3. {\em Relative difference of potential energies calculated with the
exact and interpolating formulae for 
the polarization operator for $g=0.5$, $m=0.1$.}
\end{center}
}

 The accuracy appears to be the worst at
$z\approx 1/m$. We have checked that for various values of the
ratio $g/m$, the relative accuracy of (\ref{8})  is
better than 4\%  at all $z$ .

\section{Screening of the Coulomb potential in $\boldsymbol{D=4}$ QED in a
superstrong magnetic field}

In order to write an expression for the electric potential of a
point-like charge we need a formula for the photon polarization
operator in an external magnetic field $B$. This quantity was studied
in many papers; a detailed review and references can be found in \cite{7}.
In the calculation of the polarization operator, electron propagators
in an external magnetic field are used. In the case of a strong
field $B\gg B_{cr} = m_e^2/e$ the spacing of Landau levels  is much larger
than the electron mass. This is why the contribution of the electrons
from LLL obeying the dispersion law $\varepsilon_0^2 = m_e^2 +
p_z^2$ dominates in the electron propagator. A linear dependence of
the polarization operator on the external field $B$ follows. Finally, the
electric potential writes \cite{1,3}:
\begin{equation}
{\bf\Phi}(k) =  \frac{4\pi e}{k_\parallel^2 + k_\bot^2 + \frac{2
e^3 B}{\pi} {\rm exp}\left(-\frac{k_\bot^2}{2eB}\right)
P\left(\frac{k_\parallel^2}{4m_e^2}\right)} \; , \label{11}
\end{equation}
where $k = (0, k_x, k_y, k_z)$, $k_\bot^2 = k_x^2 + k_y^2$, $k_z =
k_\parallel$ and the magnetic field $B$ is directed along the $z$
axis. As  stated in the introduction  in (\ref{11})  a quasi
two-dimensional formula for the photon polarization operator
occurs. It was obtained in \cite{2}. Expression (\ref{11}) is valid for
strong magnetic fields $B\gg B_{cr} = m_e^2/e$ and for longitudinal
momenta $k_\parallel^2 \ll eB$ (the motion of the virtual electrons
contributing to the photon polarization operator $\Pi$ occurs
in one space and one time dimensions only for $z$ larger than the Landau
radius $a_H = 1/\sqrt{eB}$; for $z < a_H$ the motion is four
dimensional).

Plugging in the Fourier transform of (\ref{11}) the interpolating formula  
(\ref{7})  for $P$
introduced in Section 2 for the electric potential of a point-like charge
along the direction of magnetic field
we get:
\begin{eqnarray}
{\bf\Phi}(z) & = & 4\pi e \int\frac{e^{ik_\parallel z} d
k_\parallel d^2 k_\bot/(2\pi)^3}{k_\parallel^2 + k_\bot^2 +
\displaystyle\frac{2 e^3B}{\pi} {\rm
exp}(-k_\bot^2/(2eB))(k_\parallel^2/2m_e^2)/(3+k_\parallel^2/2m_e^2)}
= \nonumber \\
& = & \frac{2e}{2\pi|z|} \int\limits^\infty_0 \cos(k_\parallel z)
\ln\left[ \frac{\Lambda^2 + k_\parallel^2 + \frac{2e^3B}{\pi}
\frac{k_\parallel^2}{k_\parallel^2 + 6m_e^2}}{k_\parallel^2 +
\frac{2e^3B}{\pi} \frac{k_\parallel^2}{k_\parallel^2 +
6m_e^2}}\right] d(k_\parallel|z|) \;\; ,
 \label{12}
\end{eqnarray}
where $\Lambda$ is an ultraviolet cutoff in $k_\bot$ and as we will see later
the values
 $(k_\bot^2)_{max} \equiv \Lambda^2 \la 2 eB$ are essential.
This condition allows us to replace the exponent in the denominator of the first
line of (\ref{12}) by one.  This is why this exponent
is absent in the second line of (\ref{12}). Integrating by parts we get that, 
indeed,
$\Lambda^2 \la k_\parallel^2$ is important, and $k_\parallel^2 \ll eB$
for $z\gg a_H$. So doing we obtain:
\begin{eqnarray}
{\bf\Phi}(z) & = & \frac{2e}{\pi|z|}\int\limits^\infty_0
\frac{\sin(k_\parallel z) d(k_\parallel|z|)}{k_\parallel
z\left(1+\frac{2e^3 B/\pi}{k_\parallel^2  + 6m_e^2
}\right)} \left[1+ \frac{12e^3 B m_e^2 }{\pi(k_\parallel^2
+ 6m_e^2)^2}\right] =  \\
& = & \frac{2e}{\pi|z|}\int\limits^\infty_0 \frac{\sin(k_\parallel
z)}{(k_\parallel z)} d(k_\parallel |z|) \left[1+ \frac{6m_e^2
}{k_\parallel^2  + 6m_e^2 } - \frac{2e^3 B/\pi +
6m_e^2 }{k_\parallel^2  + 2e^3 B/\pi + 6m_e^2 }
\right] =
\nonumber \\
& = & \frac{2e}{\pi|z|}\left\{\frac{\pi}{2} + \frac{1}{2i}
\int\limits^\infty_{-\infty} \frac{e^{ik_\parallel z}}{k_\parallel
z} d(k_\parallel|z|) \left[\frac{6m_e^2 }{k_\parallel^2  +
6m_e^2} - \frac{2e^3 B/\pi + 6m_e^2 }{k_\parallel^2
+ 2e^3 B/\pi + 6m_e^2 } \right]\right\} \;\; . \nonumber
\label{13}
\end{eqnarray}

Performing the last integrals with the help of residues we finally
obtain:
\begin{equation}
{\bf\Phi}(z) = \frac{e}{|z|}\left[ 1-e^{-\sqrt{6m_e^2}\;|z|} +
e^{-\sqrt{(2/\pi) e^3 B + 6m_e^2}\;|z|}\right] \;\; . \label{14}
\end{equation}

For magnetic fields $B \ll 3\pi m_e^2/e^3$ the potential is Coulomb
up to small power suppressed terms:
\begin{equation}
{\bf\Phi}(z)\left| \begin{array}{l}
~~  \\
e^3 B \ll 3\pi m_e^2
\end{array}
\right. = \frac{e}{|z|}\left[ 1+ O\left(\frac{e^3
B}{3\pi m_e^2}\right)\right] \label{15}
\end{equation}
in full agreement with the $D=2$ case  (see (\ref{9}), where $g^2$
plays the role of $e^3 B$). In the opposite case of superstrong
magnetic fields $B\gg 3\pi m_e^2/e^3$ we get:
\begin{equation}
{\bf\Phi}(z) = \left\{
\begin{array}{lll}
\frac{e}{|z|} e^{(-\sqrt{(2/\pi) e^3 B}\;|z|)} & , & |z| <
\frac{1}{\sqrt{(2/\pi) e^3 B}}\ln\left(\sqrt{\frac{e^3 B}{3\pi
m_e^2}}\right) \\
\frac{e}{|z|}(1- e^{(-\sqrt{6m_e^2}\;|z|)}) & , & \frac{1}{m_e} > |z| >
\frac{1}{\sqrt{(2/\pi)e^3 B}}\ln\left(\sqrt{\frac{e^3 B}{3\pi
m_e^2}}\right) \\
\frac{e}{|z|} & , & |z| > \frac{1}{m_e}
\end{array}
\right. \;\; , \label{16}
\end{equation}
which also closely resembles $D=2$ case, (\ref{10}), with the
substitution $e^3 B \to g^2$.

The expression for the potential energy of two opposite charges separated by
the distance $z$ along the magnetic field (the transverse separation is
zero) is:
\begin{equation}
\bar V(z) = - e{\bf\Phi}(z) = -\frac{e^2}{|z|} \left[
1-e^{-\sqrt{6m_e^2}\;|z|} + e^{-\sqrt{(2/\pi) e^3 B +
6m_e^2}\;|z|}\right] \;\; . \label{17}
\end{equation}

In Fig. 4 the potential energy which follows from eq. (\ref{17})
for  a magnetic field $B=10^{17}$ G is shown as well as its
asymptotics at large distances:
\begin{equation}
\bar V(z) = -\frac{e^2}{|z|} \;\; ,\;\; |z| > \frac{1}{m_e} \;\; ,
\label{18}
\end{equation}
and at small distances:
\begin{equation}
\bar V(z) = -\frac{e^2}{|z|} e^{-\sqrt{(2/\pi)e^3 B}\;|z|} \;\; ,\;\; z <
\frac{1}{\sqrt{(2/\pi)e^3 B}} \ln\sqrt{\frac{e^3 B}{3\pi m_e^2}}
\;\; . \label{19}
\end{equation}

\vbox{
\begin{center}
\bigskip
\includegraphics[width=.7\textwidth]{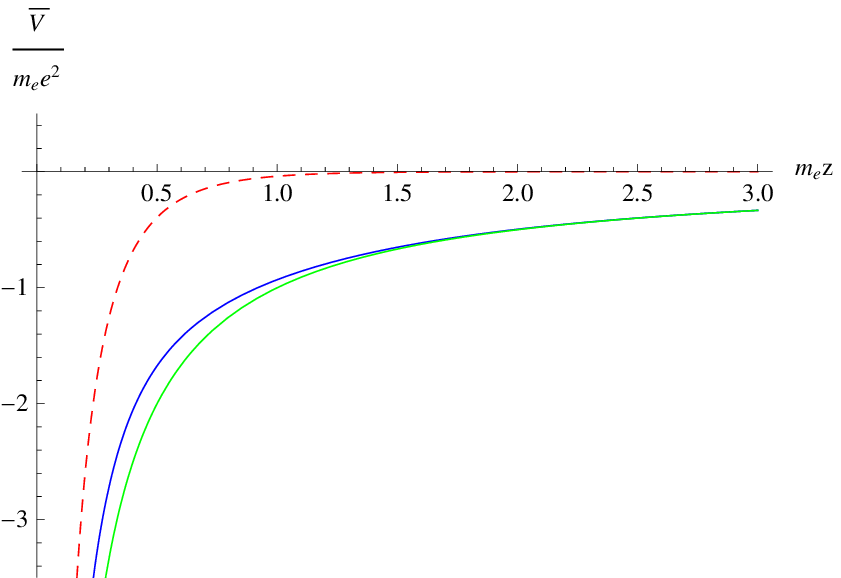}

\medskip
Fig. 4. {\em Modified Coulomb potential energy at $B=10^{17}$G (blue) and its
long distance (green-pale) and short distance (red-dashed) asymptotics.}

\end{center}
}

In the following two figures (Fig. 5 and Fig. 6)
the analytical formula (\ref{17})   for the
potential energy is compared with the result of a numerical integration
using the exact formula for $P(k_\parallel^2/4m_e^2)$ given by
equation (\ref{3}). The same value of the magnetic field
$B=10^{17}$ G is used. We checked that for various values of $B$
the agreement between analytical and numerical integrations is
never worse than $3$\%.

\vbox{
\begin{center}
\bigskip
\includegraphics[width=.7\textwidth]{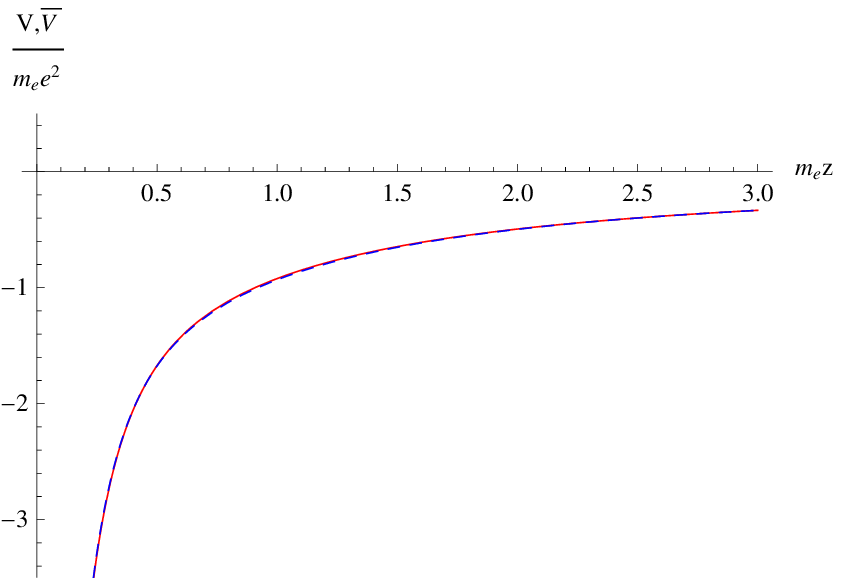}

\medskip
Fig. 5. {\em Modified Coulomb potential energy at $B=10^{17}$G.
Analytical formula (approximate $\bar P$) in blue-dashed versus numerical
integration (exact $P$) in red (the two curves practically coincide).}

\end{center}
}

\vbox{
\begin{center}
\bigskip
\includegraphics[width=.7\textwidth]{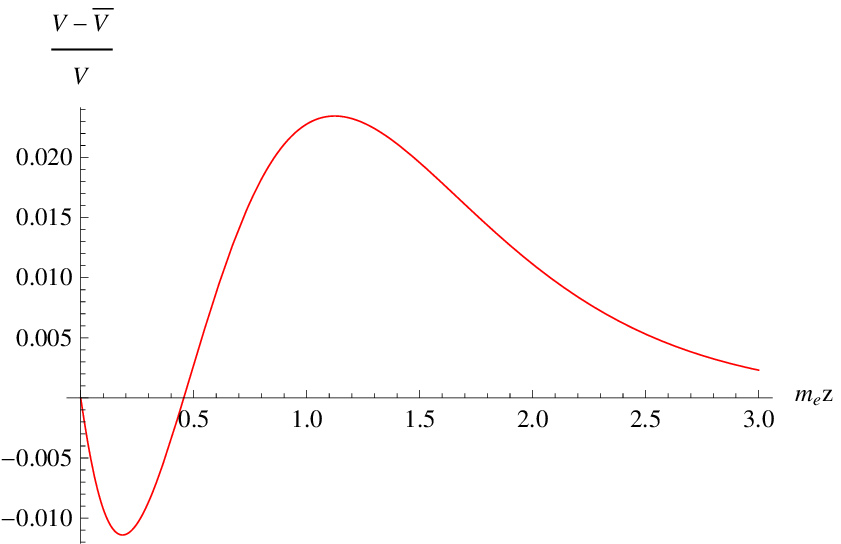}

\medskip
Fig. 6. {\em Relative accuracy of the analytical formula for the 
modified Coulomb potential
energy at $B=10^{17} $G.}

\end{center}
}

In the next section we will use the analytical expression for the
 electric potential energy (\ref{17}) to determine the energies of a
hydrogen atom in superstrong magnetic fields.

\section{Energies of hydrogen atomic levels originating from LLL in
superstrong magnetic fields}

\subsection{General considerations}

We are interested in the spectrum of a hydrogen atom in a
superstrong magnetic field $B$. In the absence of magnetic field
the spatial size of the wave function of the ground state atomic
electron is
characterized by the Bohr radius $a_B = 1/(m_e e^2)$, its energy
equals $E_0 = -m_e e^4/2 \equiv -Ry$, where $Ry$ is the Rydberg
constant. The transverse (with respect to $B$)
size of the ground state of the electron wave
function in an external magnetic field $B$ is characterized
by the Landau radius $a_H =
1/\sqrt{eB}$. The Larmour frequency of the electron precession is
$\omega_L = eB/m_e$. For a  magnetic field $B_a = e^3 m_e^2 = 2.35
\cdot 10^9$ G called ``atomic magnetic field'',
these sizes and energies are close to
each other: $a_B = a_H$, $E_0 \sim \omega_L$. We wish to study the
spectrum of the hydrogen atom in magnetic fields much larger than
$B_a$. In this case the motion of the electron is mainly controlled
by the
magnetic field: it makes many oscillations in this field before it
makes one in the Coulomb field of the nucleus. This is the condition for
applicability of the adiabatic approximation, used for this
problem for the first time in \cite{8}.

The spectrum of a Dirac electron in a pure magnetic field is well
known \cite{9}; it admits a continuum of energy levels due to the free
motion along the field:
\begin{equation}
\varepsilon_n^2 = m_e^2 + p_z^2 + (2n+1 + \sigma_z) eB  \;\; ,
\label{20}
\end{equation}
where $n = 0, 1,2, ...$; $\sigma_z = \pm 1$ is
the spin projection of the electron on $z$ axis
multiplied  by two. For magnetic fields
larger than $B_{cr} = m_e^2/e$, the electrons are relativistic with
only one exception: electrons belonging to the lowest Landau level (LLL,
$n=0$, $\sigma_z = -1$) can be non-relativistic.

In what follows we
will study the spectrum of electrons from LLL
in the  Coulomb field of the proton  modified by the
superstrong $B$  . The solution, in cylindrical coordinates
$(\vec{\rho}, z),$  of the
Schr\"{o}dinger equation for an electron in a  magnetic field $B$
which is homogeneous and constant in
time,
in the gauge in which $\vec{A}  =
\frac{1}{2} \;\vec{B} \times \vec{r}$ can be
found in \cite{10}. The electron energies are:
\begin{equation}
E_{p_z n_\rho m \sigma_z} = \left(n_\rho + \frac{|m| +m+1+
\sigma_z}{2}\right)\frac{eB}{m_e} + \frac{p_z^2}{2m_e} \;\; ,
\label{21}
\end{equation}
where $n_\rho = 0,1,2,...$ is the number of nodal surfaces, 
$m = 0, \pm 1, \pm 2, ...$ is the electron
orbital momentum projection on the  $z$ axis (direction of the magnetic
field)
and $\sigma_z = \pm 1$.  According to \cite{10}, the LLL
wave functions are:
\begin{equation}
R_{0m}(\vec\rho) = \left[\pi(2a_H^2)^{1+|m|} (|m|!)\right]^{-1/2}
\rho^{|m|}e^{\left(im\varphi - \rho^2/(4a_H^2)\right)} \;\; , \label{22}
\end{equation}
$$
\rho = |\vec{\rho}|, \;\;\;
\int|R_{0m}(\vec\rho)|^2 d^2 \rho = 1 \; , \;\;  m=0,-1,-2,...
$$

We should now take into account the electric potential of the atomic
nucleus located at $\vec{\rho} = z = 0$. For $a_H \ll a_B$
the adiabatic
approximation can be used and the wave function
can be looked for in the following
form:
\begin{equation}
\Psi_{n 0 m (-1)} = R_{0m}(\vec{\rho})\chi_n(z) \;\; , \label{23}
\end{equation}
where $\chi_n(z)$ is the solution of a Schr\"{o}dinger equation
for an electron motion along the direction of the magnetic field:
\begin{equation}
\left[-\frac{1}{2m_e} \frac{d^2}{d z^2} + U_{eff}(z)\right]
\chi_n(z) = E_n \chi_n(z) \;\; . \label{24}
\end{equation}
Without screening the effective potential is given by the
following formula \cite{4}:
\begin{equation}
U_{eff} (z) = -e^2\int\frac{|R_{0m}(\vec{\rho})|^2}{\sqrt{\rho^2 +
z^2}}d^2 \rho \;\; , \label{25}
\end{equation}
which becomes the Coulomb potential for $|z| \gg a_H$
\begin{equation}
U_{eff}(z) \left|
\begin{array}{l}
~~  \\
z \gg a_H
\end{array}
\right. = - \frac{e^2}{|z|} \;\;  \label{26}
\end{equation}
and is regular at $z=0$

\begin{equation}
U_{eff}(0)
 \sim - \frac{e^2}{|a_H|} \;\; . \label{260}
\end{equation}

To take screening into account we must use (\ref{17}) to modify
(\ref{25}) (see below). Since $U_{eff}(z) = U_{eff}(-z)$, the wave
functions are odd or even under reflection $z\to -z$; the ground
states (for $m=0, -1, -2, ...$) are described by even wave
functions.

In Fig.7 the different scales important in the consideration of the hydrogen 
atom in strong magnetic field are shown.

\vbox{
\begin{center}
\bigskip
\includegraphics[width=.7\textwidth]{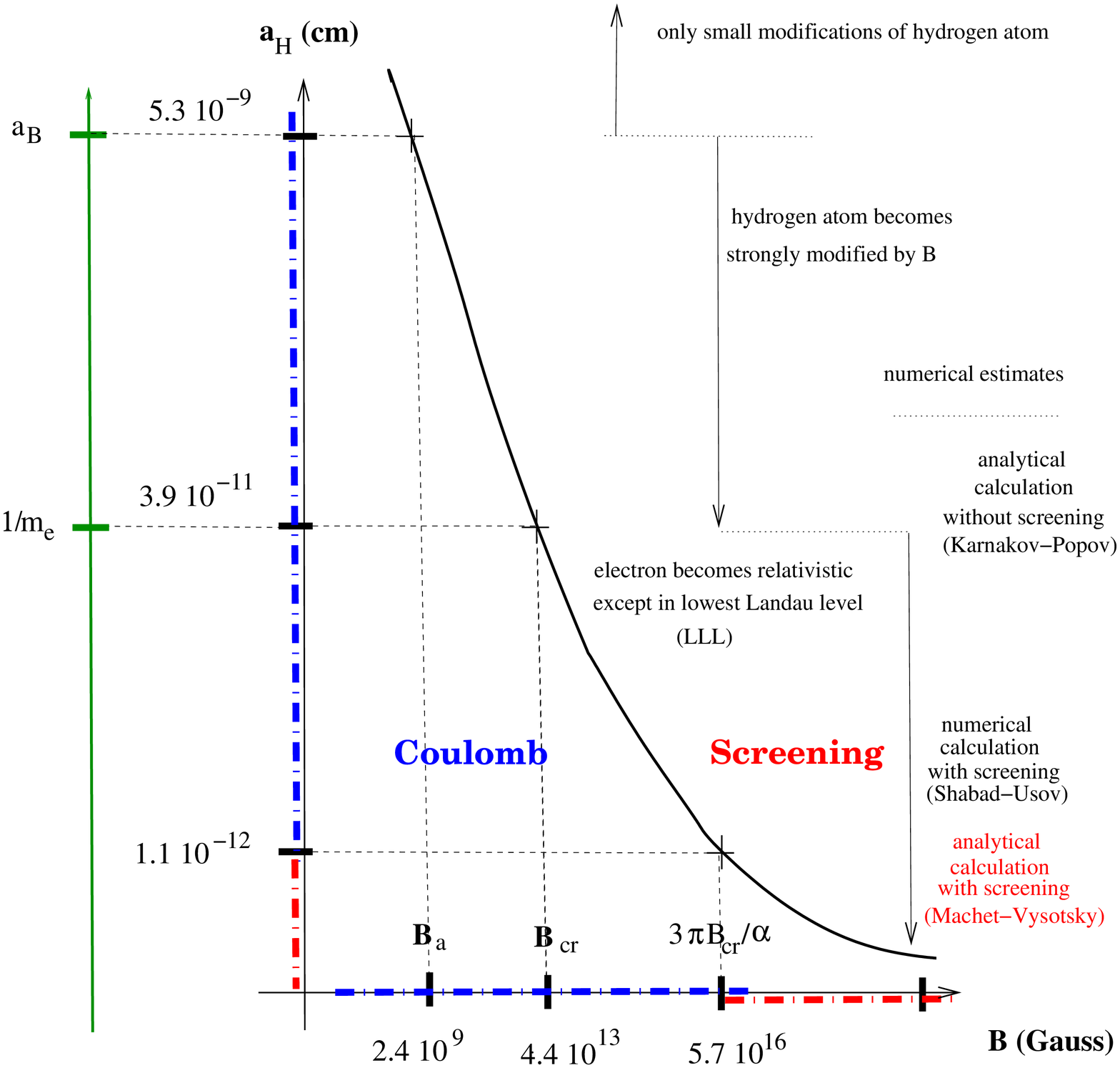}

\medskip
Fig. 7. {\em Landau radius $a_H$ versus magnetic field $B$.}

\end{center}
}

\bigskip

\subsection{The shallow-well approximation}

This approximation was used in \cite{1} for the calculation
of the ground state energy of the
hydrogen atom  in a superstrong magnetic field, taking
screening into account. The authors of \cite{1} follow \cite{5},
where this approximation was used for calculating  the ground
state energy of hydrogen atom  without screening. As  pointed
out in \cite{4}, the accuracy of the shallow-well approximation
for the
problem under consideration is very poor. Without taking the
screening into account it can be estimated as:
\begin{equation}
\frac{\delta E}{E} = \left(\frac{1}{1-\displaystyle\frac{\ln(\ln^2 H)}{\ln
H}}\right)^2 -1 \;\; , \label{27}
\end{equation}
where $H$ is a magnetic field measured in units of the atomic
field $B_a$, $H \equiv B/B_a = (a_B/a_H)^2$. Even for $B = B_{cr} = m_e^2/e = 4.41
\cdot 10^{13}$ G ($H\approx 2 \cdot 10^4$) we obtain $\delta E/E \approx
3.5$, while for $B = 10^{17}$ G ($H = 5 \cdot 10^7$) $\delta E/E
\approx 1.4$.

However the formula for the ground state energy in a shallow-well
approximation taken from \cite{5} and used in \cite{1}:
\begin{equation}
E_{\cite{5}}^{sw} = -2m_e\left[\int\limits_{a_H}^{a_B} U(z) dz\right]^2
\label{28}
\end{equation}
allows us to check with which accuracy the integral of our
expression (\ref{17}) reproduces the result obtained in \cite{1},
where the potential $A_0(z)$ found numerically was integrated.
According to eq. (14) from P.R.L. paper \cite{1}
\begin{equation}
E_{\rm lim}^{sw} = - 2m_e\alpha^2 \times 73.8 = - 4 \; \mbox{\rm keV} \;\;
, \label{29}
\end{equation}
where $E_{\rm lim}$ is the limiting value of the ground state
energy of a
 hydrogen atom at $B \to \infty$. Substituting
(\ref{17}) in (\ref{28}) we get:
\begin{eqnarray}
E_{\rm analytical}^{sw}
& = &- 2m_e e^4 \left[\ln\left(\frac{a_B}{a_H}\right) + \ln
(\sqrt{6m_e^2}\;a_H) - \ln(\sqrt{6m_e^2 + \frac{2}{\pi}e^3 B}\; a_H)
\right]^2 = \nonumber \\
& = & -2m_e e^4 \ln^2 \sqrt{\frac{H}{1+\displaystyle\frac{e^6}{3\pi} H}} 
\label{30}
\end{eqnarray}
and in the limit $H \gg 3\pi/e^6$

\begin{eqnarray}
E^{sw}_{analytical}\Bigg|_{H\rightarrow \infty} = 
-2m_e e^4 \ln^2 \left(\sqrt{\frac{3\pi}{e^6}}\right) =
- 2m_e \alpha^2 \times 72.3 \;\; , \label{300}
\end{eqnarray}
which coincides with (\ref{29}) within 2\% accuracy
($e^2 \equiv \alpha$ is
used). It means that integrals over potential energy differ by 1\%
only.

Let us present the derivation of the formula for the ground state
energy of a particle with mass $\mu$ undergoing a one-dimensional 
motion in a
shallow-well potential \cite{39}. One starts from the Schr\"{o}dinger
equation:
\begin{equation}
-\frac{1}{2\mu} \frac{d^2}{dz^2} \chi(z) + U(z) \chi(z) = E_0
\chi(z) \;\; . \label{31}
\end{equation}
Neglecting $E_0$ in comparison with $U$ and integrating (\ref{31})
we get:
\begin{equation}
\chi^\prime(a) = 2\mu\int\limits_0^a U(x)\chi(x) dx \;\; ,
\label{32}
\end{equation}
where we assume that $U(x) = U(-x)$ (this is why $\chi(x)$ is an
even  function). The next assumptions are: 1. the finite range of
the potential
energy: $U(x) \neq 0$ for $a> x > -a$; 2.
$\chi$ undergoes very small variations inside the well. 
Since outside the well $\chi(x) \sim
e^{-\sqrt{2\mu |E_0|}\;x}$, we readily obtain:
\begin{equation}
|E_0| = 2\mu\left[\int\limits_0^a U(x) dx\right]^2 \;\; . \label{33}
\end{equation}

For
\begin{equation}
\mu |U| a^2 \ll 1
\label{34}
\end{equation}
(condition for the potential to form a shallow well) we get that,
indeed, $|E_0| \ll |U|$ and that the variation of $\chi$ inside the well is
small, $\Delta \chi/\chi \sim \mu |U|a^2 \ll 1$.

Concerning the one-dimensional Coulomb potential, it satisfies
the condition (\ref{34}) only for $a\ll 1/(m_e e^2)\equiv a_B$,
so (\ref{33})
cannot be used. This explains why the accuracy of (\ref{28}) is
very poor.

\bigskip

\subsection{The Karnakov--Popov equation \cite{4}}

It provides a several percent accuracy for the ground state energy
for  $H > 10^3$. It will be derived in this subsection
 and, in the
next one, we will obtain the modification of this equation
due to screening.
The Karnakov-Popov (KP) equation determines the energies of the
families of even states with
given $n_\rho$ and $m$. The starting point of its derivation
is equation (\ref{24}). Karnakov and Popov integrated it
over $z$, from $z=0$ till
a value of $z$ such that $a_H \ll z \ll a_B$, where the
assumptions of a shallow well potential ($|U| \gg |E_0|$, $m_e |U|a^2
\ll 1$) can be used. Assuming $\chi(z) \approx 1$ we get:
\begin{eqnarray}
\chi^\prime(z) & = & -2m_e e^2 \int\limits_0^z dx
\int\frac{|R_{0m}(\vec\rho)|^2}{\sqrt{\rho^2 + x^2}} d^2 \rho =
\nonumber \\
& = & -2m_e e^2 \int \ln \left(\frac{z}{\rho} +
\sqrt{\frac{z^2}{\rho^2} +1}\right) |R_{0m}(\vec\rho)|^2 d^2\rho \;\;
. \label{35}
\end{eqnarray}
Since $R_{0m}(\vec\rho)$ decreases exponentially for $\rho > a_H$ and
since
$z$ is much larger than $a_H$, expanding the logarithm in powers of
 $\rho/z$, we can
neglect the terms suppressed as $(\rho/z)^n$:
\begin{equation}
\ln\left(\frac{z}{\rho} + \sqrt{\frac{z^2}{\rho^2}+1}\right) =
\ln\left(2\frac{z}{\rho}\right) = \ln\left(\frac{z}{a_H}\right) +
\ln\left(\frac{2a_H}{\rho}\right) \;\; , \label{36}
\end{equation}
\begin{eqnarray}
\chi^\prime(z) & = & -2m_e e^2 \left[\ln\left(\frac{z}{a_H}\right)
+ \int\ln\left(\frac{2a_H}{\rho}\right)|R_{0m}(\vec\rho)|^2
d^2\rho\right]= \nonumber \\
& = & - 2m_e e^2 \ln\left(\frac{z}{a_H}\right) + A_{|m|0} \;\; ,
\label{37}
\end{eqnarray}
in which
\begin{equation}
A_{|m|0} = -m_e e^2\left[\ln(2) - \int\limits_0^\infty\ln(x)
x^{|m|} e^{-x} \frac{dx}{|m|!}\right] \;\; , \label{38}
\end{equation}
and the expression (\ref{22}) for $R_{0m}(\vec\rho)$ was used. To
calculate the last integral the following equality must be used:
\begin{equation}
\int\limits_0^\infty \ln(x) x^{|m|} e^{-x} dx =
\Gamma^\prime(|m|+1) = \Gamma(|m| +1) \psi(|m| +1) \;\; ,
\label{39}
\end{equation}
where $\psi$ is the logarithmic derivative of the gamma function.

Finally we reproduce eq. (2.11) from \cite{4}:
\begin{equation}
A_{|m|0} = -m_e e^2[\ln(2) - \psi(1+|m|)] \;\; , \label{40}
\end{equation}
where $\psi(1) = -\gamma = -0.5772$, $\gamma$ is Euler constant and
$\psi(n+1) = -\gamma + \sum_{k=1}^n(1/k)$ for $n=1,2,3,...$.

At $z \gg a_H$ the solution of the Schr\"{o}dinger equation with the
Coulomb potential which exponentially decreases at $z \gg a_B$ is
given by the Whittaker function \cite{4}.

For the state with energy
\begin{equation}
E = -(m_e e^4/2)\lambda^2 \label{41}
\end{equation}
the logarithmic derivative of Whittaker function at $z \ll a_B$ is
\cite{4}:
\begin{equation}
\frac{W^\prime}{W} = -m_e e^2 \left\{ 2\ln(z/a_B) + \lambda + 2\ln
\lambda + 2\psi\left(1-\frac{1}{\lambda}\right) + 4\gamma + 2\ln 2 +
O\Big((z/a_B)\ln\lambda\Big)\right\} \;\; . \label{42}
\end{equation}

From the condition $W^\prime/W = \chi^\prime$ we obtain an
equation for the energies of the even bound states of the hydrogen atom
originating from LLL ($n_\rho = 0$, $m=0, -1, -2, ...$):
\begin{eqnarray}
&&2\ln\left(\frac{z}{a_H}\right) + \ln 2 - \psi(1+|m|) + O(a_H/z)
= \\
&& =  2\ln\left(\frac{z}{a_B}\right) + \lambda + 2\ln \lambda +
2\psi\left(1-\frac{1}{\lambda}\right) + 4\gamma + 2\ln 2 +
O(z/a_B) \;\; , \nonumber \label{43}
\end{eqnarray}
and for $a_B \gg a_H$, choosing $a_B \gg z \gg a_H$, we may neglect
non-calculated terms. This leads to the KP equation ( \cite{4}, (2.11, 2.13)):
\begin{equation}
\ln(H) = \lambda + 2\ln\lambda +
2\psi\left(1-\frac{1}{\lambda}\right) + \ln 2 + 4\gamma +
\psi(1+|m|) \;\; , \label{44}
\end{equation}
where we recall that
$H \equiv B/B_a = B/(e^3m^2_e) $.
We postpone the analysis of this eigenvalue equation till the next
subsection, where its modification due to screening will be
established.

The energies of the odd states are \cite{4}:
\begin{equation}
E_{\rm odd} = -\frac{m_e e^4}{2n^2} + O\left(\frac{m_e^2
e^3}{B}\right) \; , \;\; n = 1,2, ... \;\; . \label{45}
\end{equation}
So, for superstrong magnetic fields $B \sim m_e^2/e^3$ the deviations of
odd states
from the Balmer series are negligible.

\bigskip

\subsection{ Equation for the energies of even states with screening}

We must equate the logarithmic derivative of wave function obtained
by integrating the Schr\"{o}dinger equation at small distances with the
logarithmic derivative of the Whittaker function. To do this we should
choose $z$ to be larger than the distance at which the Coulomb
potential is screened. That is why the condition $z \gg a_H$ which
was used in the case of the absence of screening is not enough.
For $z_0 > 1.5/m_e$ the corrections coming from screening in (\ref{17})
are only a few percent. The wave function at $z\ga z_0$ is thus
described  by the
Whittaker function with good accuracy.

To derive an analog of (\ref{43}), it is necessary that $\chi$ stays
practically constant from
$z=0$ up to the domain where the effective potential becomes
Coulombian with high
accuracy, and, accordingly, where the Whittaker function provides
a very good approximation to the  wave function.
The characteristic distance along which
the ground state wave function varies in the $z$ direction is,
according to
the uncertainty relation:
\begin{equation}
z_{\rm char} \sim \frac{1}{\sqrt{2m_e(-E)}} \sim a_B/\lambda \;\; ,
\label{46}
\end{equation}
and it decreases when $B$ increases since $\lambda \sim \ln B$
(see eq. (\ref{44})).
 However,
screening helps: with the account of it $\lambda$
goes to a finite value, $\lambda_\infty 
\approx 11$ at $B \to\infty$  (see below).
 For the size of a ground state wave
function we get:
\begin{equation}
z_\infty \sim \frac{137}{11 m_e} \approx \frac{12}{m_e} \;\; ,
\label{47}
\end{equation}
and for
\begin{equation}
1.5/m_e < z_0 < 3/m_e \label{48}
\end{equation}
the potential (\ref{17}) is Coulombian with good accuracy while the
variation of the wave function $[\chi(0) - \chi(z_0)]/\chi(0)$ is
small.

When screening is taken into account an expression for effective 
potential (\ref{25}) transforms into

\begin{equation}
\tilde U_{eff} (z) = -e^2\int  \frac{|R_{0m}(\vec{\rho})|^2}
{\sqrt{\rho^2 +z^2}} d^2\rho
\left[1-e^{-\sqrt{6m_e^2}\;z}  
+  e^{-\sqrt{(2/\pi)e^3 B + 6m_e^2}\;z}\right] 
 \;\; , \label{520}
\end{equation}
 see Fig. 8. On the same figure a simplified potential given by 
\begin {equation}
U_{simpl} (z) =  -e^2 \frac{1}{\sqrt{a_H^2 +z^2}}
\left[1-e^{-\sqrt{6m_e^2}\;z}  
+ e^{-\sqrt{(2/\pi)e^3 B + 6m_e^2}\;z}\right]
 \;\; , \label{530}
\end{equation}
is shown.

\vbox{
\begin{center}
\bigskip
\includegraphics[width=.7\textwidth]{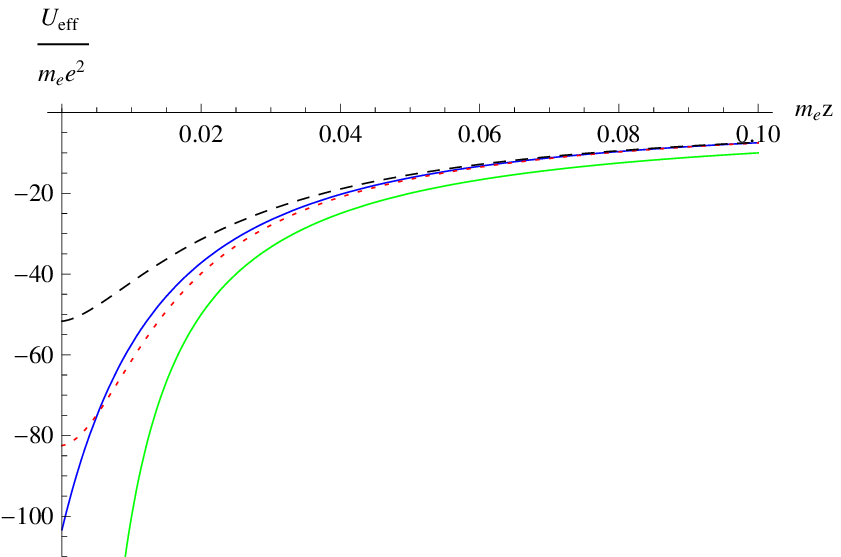}

\medskip
Fig. 8. {\em Effective potential with screening for $m=0$ (blue)
and  $m=-1$ (black-dashed), (\ref{520}); simplified potential 
(\ref{530}) (red-dotted).
The curves correspond to $B=3 \cdotp 10^{17}$ G. Coulomb potential 
(green-pale) is also shown.}

\end{center}
}

The  screening  changes
(\ref{35}) into:
\begin{eqnarray}
\tilde\chi^\prime(z_0) = -2m_e e^2 \int\limits_0^{z_0} dx
\int\frac{|R_{0m}(\vec\rho)|^2}{\sqrt{\rho^2 + x^2}}
\left[1-e^{-\sqrt{6m_e^2}\;x} + \right. \nonumber \\
+ \left. e^{-\sqrt{(2/\pi)e^3 B + 6m_e^2}\;x}\right] d^2\rho \equiv
\chi^\prime(z_0) + \Delta \chi^\prime(z_0) \;\; . \label{49}
\end{eqnarray}
Calculating  integrals with exponents we get:
\begin{eqnarray}
\int_0^{z_0}\frac{dx}{\sqrt{\rho^2 + x^2}}e^{-ax} & \equiv &
\int_0^{z_0 a} \frac{dy}{\sqrt{y^2 + \rho^2 a^2}}e^{-y} \approx
\int_0^\infty\frac{dy}{\sqrt{y^2 + \rho^2 a^2}}e^{-y}\approx  \nonumber \\
& \approx & \ln\left(\frac{1}{\rho a}\right) + {\rm const} \;\; ,
\label{50}
\end{eqnarray}
where, in the first approximate equality, we took into account that
the integrand is exponentially damped for $y>1$, and
$z_0\sqrt{6m_e^2} \ga 4$;  in the second approximate equality
we took into account that $\rho a \approx a_H\sqrt{6m_e^2
+(2/\pi)e^3 B} \ll 1$ (because $R_{0m}$ is exponentially damped for
$\rho > a_H$ and, thus, only the values $\rho \sim a_H$ are important).

This leads to:
\begin{equation}
\Delta\chi^\prime(z_0) = -2m_e e^2 \ln\sqrt{\frac{6m_e^2}{(2/\pi)
e^3 B + 6m_e^2}} \;\; . \label{51}
\end{equation}

Equation (\ref{17}) takes into account the modification of the
Coulomb potential at zero transverse direction, $\rho =0$. In
(\ref{49}) the modified potential at $\rho \neq 0$ should be used.
We changed the factor $1/|z|$ occurring in eq. (\ref{17})
into $1/\sqrt{\rho^2 + x^2}$ in (\ref{49}).
However, according to numerical calculations
of the electric potential in the direction transverse to
that of the magnetic field
(at $z=0$) made in \cite{1}, the screening occurs at $\rho \neq 0$ as
well. To estimate the role of the  `` transverse screening'',
we substituted  $\sqrt{\rho^2 + x^2}$ for $x$ inside the exponents of (\ref{49}).
The corrections arising in eq. (\ref{50}) turn out to be of the order of
$\rho a \la\sqrt{e^3 B/eB} = e$. The resulting correction to the
ground state energy is negligible; the corrections to the
energies of the excited states are
small. In Fig. 9 we compare the effective potential given by (\ref{520})
with the one obtained by substituting in the exponents of (\ref{520})
$\sqrt{\rho^2+z^2}$ for $z$ for $m=0$ and $m = -1$.

\vbox{
\begin{center}
\bigskip
\includegraphics[width=.7\textwidth]{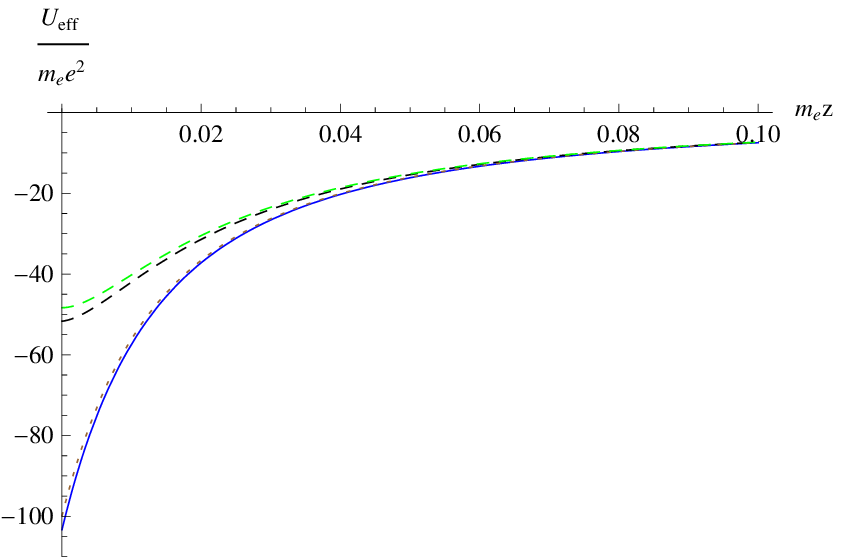}

\medskip
Fig. 9. {\em Effective potentials according to (\ref{520})
 with the replacement of $z$ by $\sqrt{\rho^2+z^2}$ in the exponents
for $m=0$ (brown-dotted)
and  $m=-1$ (green(pale)-dashed). For the comparison curves 
which directly correspond to (\ref{520}) are also shown;
 $m=0$ (blue) and  $m=-1$ (black-dashed).
The curves correspond to $B=3 \cdotp 10^{17}$ G.}

\end{center}
}

With the help of (\ref{51}) we get the modified KP equation,
which takes screening into account:
\begin{equation}
\ln\left(\frac{H}{1+\displaystyle\frac{e^6}{3\pi}H}\right) = \lambda +
2\ln\lambda + 2\psi\left(1-\frac{1}{\lambda}\right) + \ln 2 +
4\gamma + \psi(1+|m|) \;\; . \label{52}
\end{equation}
The ground state has $m=0$; for $H=10^3$,
its energy as given by (\ref{52})
coincides with the result of the numerical solution of the
Schr\"{o}dinger equation at a few percent level.

For $H\geq 10^5 $ the expansion $\psi(1-\frac{1}{\lambda}) = \psi(1) -
\frac{\pi^2}{6\lambda}$  has better than 1\%
accuracy \cite{4}. Using this approximation, we obtain:
\begin{equation}
H = 10^5 \; ; \; B = 2.3 \cdot 10^{14} \; {\rm G} \; ; \; \lambda
= 6.91 \; ; \; E_0 = -\frac{m_e e^4}{2}\lambda^2 = -0.65 \; {\rm keV}
\; . \label{53}
\end{equation}
The accuracy of the adiabatic approximation is characterized by
the ratio $a_H/a_B \equiv \sqrt{B_a/B}$ which equals 0.0032 for
$H = 10^5$.

The screening starts to be important at $H \sim 3\pi/e^6 \sim
10^7$. For example, $H = 10^7$ and $H = 1.7 \cdot 10^7$
correspond to the same ground state energy respectively
without and with screening. For $H\gg 10^7$, which corresponds to $B
\gg 10^{16}$ G, the ground state energy reaches the limiting
value
\begin{equation}
\lambda^{\rm lim} \Big|
_{B\gg 10^{16} \; {\rm G}}
  \approx 11.2 \; , \;\; E^{\rm lim}_0 \approx -1.7 \; {\rm
keV} \;\; , \label{54}
\end{equation}
which differs from the result obtained in \cite{1} by a factor 2.5.
So, when screening is accounted for, the accuracy of the shallow-well
approximation is never  better than 250\%.

For each value of $m$ ($m=0, -1, -2, ...$) eq. (\ref{52}) gives a
tower of even states, starting from the ground state, the energy
of which is very large on the Rydberg scale. Let us rewrite
(\ref{52}) for the $m=0$ tower:
\begin{equation}
\ln\left(\frac{H}{1+\displaystyle\frac{e^6}{3\pi} H}\right) = \lambda +
2\ln\lambda + 2\psi(1-\frac{1}{\lambda}) + \ln 2 + 3\gamma \;\; .
\label{55}
\end{equation}

The ground state energy corresponds to a large value of
$\lambda$, of the order
of the left hand side of (\ref{55}) as just discussed (see
(\ref{53}), (\ref{54})). $\psi(x)$ has poles at $x=0, -1, -2, ...$
which correspond to $\lambda=1, 1/2, 1/3, ...$ The Balmer
series of energies $E_n = -m_e e^4/(2n^2)$, $n=1,2,...$ that
they form would correspond to an
infinite value for the left hand side of (\ref{55}). That it
is always finite shows that even states get
shifted from their Balmer values.

Let us determine the energy of the first excited even state.
From the well-known expansion of the gamma
function at  small values of its argument we obtain the corresponding
expansion
for $\psi$:
\begin{equation}
\Gamma(\varepsilon) = \frac{1}{\varepsilon} -\gamma +
O(\varepsilon) \; , \;\; \psi(-\varepsilon) =
\frac{1}{\varepsilon}-\gamma \;\; . \label{56}
\end{equation}
Plugging in (\ref{55}) $\lambda = 1-\varepsilon$ we obtain:
\begin{equation}
\lambda =
1-\frac{2}{\ln\left(\displaystyle\frac{H}{1+\displaystyle\frac{e^6}{3\pi}H}\right)-\ln2 -
\gamma +1} \;\; . \label{57}
\end{equation}
Eq. (\ref{57}) describes how the first excited even state (the
second even state) moves towards its Balmer value $\lambda =1$
with growing $H$. When $H \to\infty$ we obtain
$\lambda_{(\infty)}^2 = 0.78$ which determines the energy of the
first excited even state in the limit of infinitely large $B$
according to the formula $E=-(m_e e^4/2)\lambda^2$. The first odd
state has the energy $E_{\rm odd}^1 = -m_e e^4/2$.

For arbitrary $n$ the analog of (\ref{57}) looks like
\begin{equation}
\lambda_n = \frac{1}{n}
-\frac{2/n^2}{\ln\left(\displaystyle\frac{H}{1+
\displaystyle\frac{e^6}{3\pi}H}\right)
 - \ln
2 - \gamma + 1/n + 2\ln n - 2\sum_{k=1}^{n-1}\frac{1}{k}} \;\; ,
\label{58}
\end{equation}
where the last sum in the denominator is the harmonic number $H_{n-1}$.

A similar formula with no screening and in the limit
$H \gg 1$ can be found in \cite{KR};
we however disagree concerning the factor $2$ in the numerator of their
eq.~(11). An expression which coincides with the leading term in (\ref{58})
for $H \gg 1$ without screening is contained in the book by Khriplovich
``Teoreticheskii kaleidoskop'', Novosibirsk, 2007, eq.~(3.39).

The energy spectrum of the hydrogen atom in the limit of infinite magnetic 
field is shown in Fig. 10.

\vbox{
\begin{center}
\bigskip
\includegraphics[width=.6\textwidth]{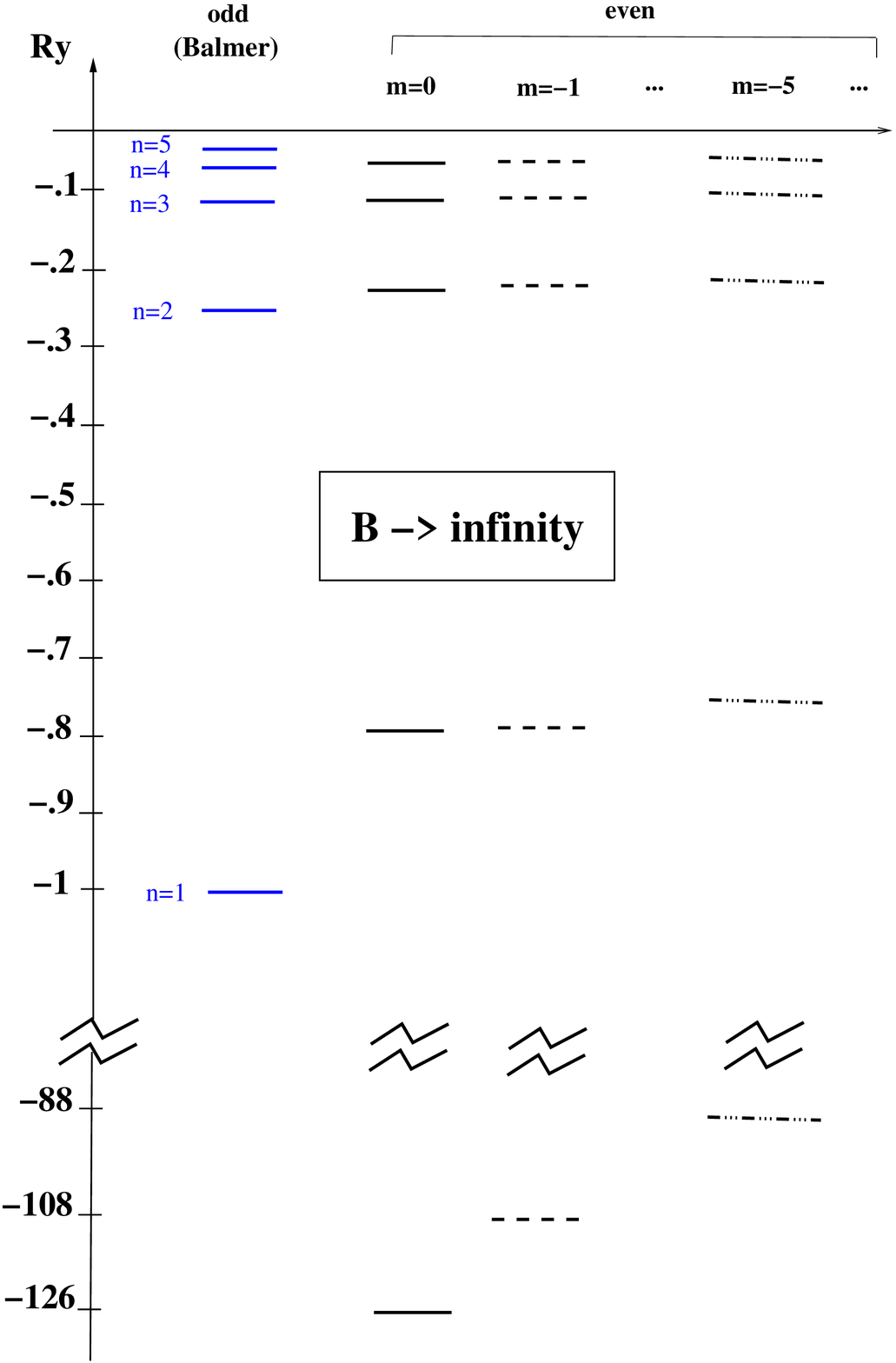}

\medskip
Fig. 10. {\em Spectrum of hydrogen levels in the limit of infinite magnetic field.
Energies are given in rydberg units, $Ry \equiv 13.6 \; eV$.}

\end{center}
}

\section{Conclusion}

An analytical formula for the Coulomb potential $\Phi(z)$ in a
superstrong magnetic field has been derived. It reproduces the results
of the numerical calculations made in \cite{1} with good accuracy.
Using it, an algebraic formula for the energy spectrum of
the levels of a hydrogen atom
originating from the lowest Landau level in a superstrong $B$ has
been obtained. The
energies start to deviate from those obtained without taking
the screening of the Coulomb potential into account \cite{4} at $B\ga
3\pi m_e^2/e^3 \approx 6 \cdot 10^{16}$ G which is much
larger than the largest magnetic field known to exist today, that
of neutron stars called magnetars (where $B
\approx 2 \cdot 10^{15}$ G may exist). Without the presented
explicit calculations we would not know at which $B$ substantial
deviations of the energies of atomic levels from those obtained without
taking screening into account take place.

\bigskip 
{\em Acknowledgments}: \;
 We are grateful to V.S. Popov for important discussions and to
S.I. Blinnikov, B.O. Kerbikov, L.B. Okun and A.V. Smilga for useful comments.

B.M. wants to thank ITEP for hospitality provided to him in June
2010 where this work was started;
M.V. is grateful to LPTHE, where this work was finished, for
hospitality provided to him in November 2010
and to grants RFBR 08-02-00494, NSh-4172.2010.2 and to
the contract of the RF Ministry of Science and Education
02.740.11.5158 for partial support.

\vspace{1cm}

\setcounter{equation}{0}
\renewcommand{\theequation}{A\arabic{equation}}

{\bf\large \noindent Appendix} \\

\noindent{\bf\large Electron mass in a strong magnetic field }

\bigskip

In the non-relativistic approximation, from (\ref{20}) we obtain:
\begin{equation}
\varepsilon_n = m_e + \frac{p_z^2}{2m_e} + (n+\frac{1}{2})
\frac{eB}{m_e} + \sigma_z \frac{eB}{2m_e} \;\; , \label{A1}
\end{equation}
where the last term originates from the interaction
of the electron magnetic moment
with the magnetic field. Taking into account the
anomalous magnetic moment of electron for the energy of the lowest
Landau level ($n=0$,  $\sigma_z = -1$) we get:
\begin{equation}
\varepsilon_0 = m_e + \frac{p_z^2}{2m} - \frac{\alpha}{4\pi}
\frac{eB}{m_e} \;\; . \label{A2}
\end{equation}
This result is valid for a magnetic field $B\la m_e^2/e$, while for
$B\gg m_e^2/e$ the correction to the energy changes sign and the
strong power dependence on $B$ is replaced by a double logarithmic one
\cite{TJ}:
\begin{equation}
\varepsilon_0 = m_e + \frac{p_z^2}{2m_e} + \frac{\alpha}{4\pi} m_e
\ln^2(eB/m_e^2) \;\; . \label{A3}
\end{equation}

As a result, for the energy of the ground state of the hydrogen
atom, we obtain:
\begin{equation}
E_0 = m_e\left[1-\frac{e^4}{2}\lambda^2 +
\frac{e^2}{4\pi}\ln^2\frac{eB}{m_e^2}\right] \;\; . \label{A4}
\end{equation}
At $H \approx 10^5$ (see (\ref{53})) the last term in brackets
compensates the second term and $E_0$ becomes
bigger than $m_e$ for a bigger $H$.
 However in the differences of the energies of
atomic states originating from the LLL the universal correction
(the last term in brackets in (\ref{A4})) cancels\footnote{We are
grateful to V.S. Popov for this remark.} and for the ``experimentally
observable'' energies of transitions between these states, the formulas
obtained in Section 4 should be used.

Concerning the double logarithmic term in (\ref{A3}) it was noted in
\cite{LS} that the appearance of an ``effective photon mass'' at
fields $B\geq m_e^2/e^3$ leads to the transformation of the double
logarithmic dependence on $B$ into a single logarithmic dependence
(concerning extraction and summation of higher 
order terms in $\alpha \;\ln B$ see also \cite{GS}, \cite{KM}).

\end{document}